\documentclass[12pt,prl,aps,superscriptaddress,tightenlines]{revtex4}

\usepackage{verbatim} 
\usepackage{graphicx}
\usepackage{dcolumn}
\usepackage{bm}
\usepackage{array}
\usepackage{booktabs}
\usepackage{amsmath,amssymb}
\usepackage{color}
\usepackage[utf8]{inputenc}

\newcommand{\BR}{{\cal B}}
\newcommand{\pp}{\pi^+\pi^-}

\newcommand{\EE}{e^+e^-}

\newcommand{\psp}{\psi(3686)}
\newcommand{\psip}{\psi(3686)}

\newcommand{\jpsi}{J/\psi}

\newcommand{\x}{X(3872)}

\newcommand{\ddstbn}{D^{*0}\bar{D}^0+c.c.}

\newcommand{\ppb}{p\bar{p}}

\begin{document}

\preprint{}

\title{Determination of the absolute branching fractions of $\x$ decays}

\author{Chunhua Li}%
 \email{chunhua@lnnu.edu.cn}
\affiliation{School of Physics and Electronic Technology, 
Liaoning Normal University, Dalian 116029, China}
\author{Chang-Zheng Yuan}
 \email{yuancz@ihep.ac.cn}
\affiliation{Institute of High Energy Physics, Chinese Academy of Sciences, Beijing 100049, China}
\affiliation{University of Chinese Academy of Sciences, Beijing 100049, China}

\begin{abstract}

We report the first determination of the absolute branching fractions of 
the $\x$ decays by globally analyzing the measurements provided 
by the Belle, BaBar, BESIII, and LHCb experiments. 
The branching fractions of $\x\to \pi^+\pi^- \jpsi$ and 
$\ddstbn$ are found to be $(4.1^{+1.9}_{-1.1})\%$ 
and $(52.4^{+25.3}_{-14.3})\%$, respectively. 
The branching fractions of the decays $\x\to\gamma \jpsi$, 
$\gamma\psp$, $\omega \jpsi$, and $\pi^0 \chi_{c1}$ are also determined. 
The global fit implies that the fraction of $\x$ 
decays which are not observed in experiments  
is $(31.9_{-31.5}^{+18.1})\%$, which indicates 
that there is still a lot of room for searching for new 
decay modes of the $\x$. With the branching fraction, we determine the 
production cross section of $\EE\to \gamma\x$ at center-of-mass 
energy 4.226~GeV. 

\end{abstract}

\date{\today}

\maketitle

Observed in 2003 by the Belle experiment~\cite{intx1}, the $\x$ has been 
the most puzzling state among all the so-called charmoniumlike
${\rm XYZ}$ states for all these years~\cite{PR_review}. Being very close 
to the $\ddstbn$ mass threshold, it is supposed to have
a large $\ddstbn$ molecule component~\cite{intx2}, 
and its large production
rate in $pp$ and $\ppb$ collision experiments suggests that 
there is a charmonium state $\chi_{c1}(2P)$ core in its 
wave function~\cite{lhcpp1,lhcpp2,lhcpp3,d0}. 
The absolute branching fractions of 
observed and unknown decays of $\x$ 
could provide essential information to 
distinguish the different scenarios~\cite{absoa1,absoa2,abso1,abso2}.

During the last 16 years, there have been many experimental 
measurements on the $\x$. Its mass was 
measured in a high precision of 170~keV~\cite{pdg} and its spin-parity
and charge-conjugate parity quantum numbers were determined
to be $J^{PC}=1^{++}$~\cite{lhcbq}; however, 
all other quantities are only known 
with very large uncertainties. Its total width is known to 
be an upper limit of 1.2~MeV at the 90\% confidence level (C.L.), and only
$\ddstbn$, $\pp\jpsi$, $\omega\jpsi$, $\gamma\jpsi$, and 
$\pi^0\chi_{c1}$ decay modes are observed with more than 5
standard deviation significance~\cite{Be2011a, Ba2008a, Be2010c, Ba2008c, Ba2010d, Bs1, Bs2, Bs3, Be2011b, Ba2009b}, and the existence of the
$\gamma\psp$ decay mode is not confirmed~\cite{Bs1, Be2011b, Ba2009b, Lb1}.
None of the above 
modes have measured absolute branching fractions due to the 
fact that the total number of $\x$ events is very difficult to be 
determined either in $B$-meson decays or in an $\EE$ annihilation
experiment. As a consequence, only the product branching fractions
in $B$ decays or the product of the total cross section and the 
$\x$ decay branching fraction is reported experimentally. 

BaBar and Belle experiments tried to measure $\BR(B^+\to \x K^+)$
by reconstructing inclusive $B^+\to K^+ + \,{anything}$ and 
searching for the monochromatic $K^+$ corresponds to two-body decay
$B^+\to \x K^+$ in the center-of-mass system of the 
$B^+$-meson~\cite{Ba2005_x,Be2018d}, but no significant signals 
were observed. By improving the algorithm of signal reconstruction
and using the full available data, BaBar is able to improve the
sensitivity of the inclusive $\x$ reconstruction, and a
$3\sigma$ significance signal of $B^+\to \x K^+$ is observed, 
and the branching fraction is determined to be
$(2.1\pm 0.6\pm 0.3\pm 0.1)\times 10^{-4}$~\cite{Ba2019d},
where the three uncertainties are statistical, systematic, and 
in the quoted $\BR(B^+\to \jpsi K^+)$, respectively.
Belle found only a $1.1\sigma$ $\x$ signal, and
provided an upper limit $\BR(B^+\to \x K^+)<2.6\times 10^{-4}$
at the 90\% C.L., 
and a central value of $(1.2\pm 1.1\pm 0.1)\times 10^{-4}$,
where the uncertainties are statistical and systematic, respectively. 

Although neither the BaBar nor the Belle measurement is significant enough 
to claim observation of $B^+\to \x K^+$ signal, we do know this decay
exists since significant signals have been observed when $\x$ is
reconstructed in its exclusive decay modes such as $\pp\jpsi$ and 
$\ddstbn$ This allows us to extract the $\x$ decay branching 
fractions by using the BaBar and Belle measurements 
of $\BR(B^+\to \x K^+)$ and the other measurements 
from Belle, BaBar, BESIII, and LHCb experiments.

The measurements used in this analysis are listed in 
Table~\ref{tab:table1}. To extract 
the absolute branching fractions of $\x$ decays we 
do a least square fit by minimizing
\begin{equation}
   \chi^2(x)=\sum_{i=1}^{25}\frac{(x_i-x)^2}{\sigma^2_i},
\end{equation}
where $i$ is the index (from 1 to 25) listed in 
Table~\ref{tab:table1}, $x_i$ are the 25 measured values,  
$x$ is the corresponding values constructed with
free parameters, and $\sigma_i$ is the sum of 
the statistical and systematic uncertainties in quadrature. 

\begin{table*}[hbtp]
\centering
\caption{\label{tab:table1}
The measurements of the $\x$ decays by Belle, BaBar, BESIII, and LHCb 
experiments: the product branching fractions of $\x\to \pp\jpsi$,
$\gamma \jpsi$, $\gamma\psp$, $\ddstbn$, and $\omega \jpsi$,
the inclusive branching fraction $\BR(B\to \x K)$,
the ratios of the branching fractions 
$\BR(\x\to \gamma \jpsi,~\omega \jpsi,~\ddstbn,~\pi^0 \chi_{c1})$ to 
$\BR(\x\to \pp\jpsi)$.
The first uncertainties are statistical and the second systematic.}
\begin{tabular}{clll}
\hline\hline
Index ($i$) & Parameters & Values               & Experiments          \\\hline
      & $\x\to \pp\jpsi$ & ($\times10^{-6}$)    &                      \\\hline
1     & $B^+\to \x K^+$  & $8.61\pm0.82\pm0.52$ & Belle~\cite{Be2011a} \\
2     &                  & $8.4\pm1.5\pm0.7$    & BaBar~\cite{Ba2008a} \\
3     & $B^0\to \x K^0$  & $4.3\pm1.2\pm0.4$    & Belle~\cite{Be2011a} \\
4     &                  & $3.5\pm1.9\pm0.4$    & BaBar~\cite{Ba2008a} \\\hline
      &$\x\to\gamma\jpsi$& ($\times10^{-6}$)    &                      \\\hline
5     & $B^+\to \x K^+$  & $1.78^{+0.48}_{-0.44}\pm0.12$ & Belle~\cite{Be2011b}\\
6     &                  & $2.8\pm0.8\pm0.1$    & BaBar~\cite{Ba2009b} \\
7     & $B^0\to \x K^0$  & $1.24^{+0.76}_{-0.61}\pm0.11$ & Belle~\cite{Be2011b}\\
8     &                  & $2.6\pm1.8\pm0.2$    & BaBar~\cite{Ba2009b} \\\hline
      & $\x\to\gamma\psp$& ($\times10^{-6}$)    &                      \\\hline
9     & $B^+\to \x K^+$  & $0.83^{+1.98}_{-1.83}\pm{0.44}$& Belle~\cite{Be2011b}\\
10    &                  & $9.5\pm2.7\pm0.6$    & BaBar~\cite{Ba2009b} \\
11    & $B^0\to \x K^0$  & $1.12^{+3.57}_{-2.90}\pm0.57$ & Belle~\cite{Be2011b} \\
12    &                  & $11.4\pm5.5\pm1.0$   & BaBar~\cite{Ba2009b} \\\hline
      & $\x\to \ddstbn$  & ($\times10^{-4}$)    &                      \\\hline
13    & $B^+\to \x K^+$  & $0.77\pm0.16\pm0.10$ & Belle~\cite{Be2010c} \\
14    &                  & $1.67\pm0.36\pm0.47$ & BaBar~\cite{Ba2008c} \\
15    & $B^0\to \x K^0$  & $0.97\pm0.46\pm0.13$ & Belle~\cite{Be2010c} \\
16    &                  & $2.22\pm1.05\pm0.42$ & BaBar~\cite{Ba2008c} \\\hline
      &$\x\to\omega\jpsi$& ($\times10^{-6}$)    &                      \\\hline
17    & $B^+\to \x K^+$  & $6\pm 2\pm 1$        & BaBar~\cite{Ba2010d} \\
18    & $B^0\to \x K^0$  & $6\pm 3\pm 1$        & BaBar~\cite{Ba2010d} \\\hline
      & Ratios           &                      &                      \\\hline
			 \specialrule{0em}{1.5pt}{1.5pt}
19    & $\frac{\BR(\x\to\gamma\jpsi)}{\BR(\x\to\pp\jpsi)}$ 
                         & $0.79\pm0.28$        & BESIII~\cite{Bs1}    \\
			 \specialrule{0em}{1.5pt}{1.5pt}
20    & $\frac{\BR(\x\to \ddstbn)}{\BR(\x\to\pp\jpsi)}$
                         & $14.81\pm3.80$       & BESIII~\cite{Bs1}    \\
			 \specialrule{0em}{1.5pt}{1.5pt}
21    & $\frac{\BR(\x\to \omega \jpsi)}{\BR(\x\to\pp\jpsi)}$
                    & $1.6^{+0.4}_{-0.3}\pm0.2$ & BESIII~\cite{Bs2}    \\
			 \specialrule{0em}{1.5pt}{1.5pt}
22    & $\frac{\BR(\x\to\pi^0 \chi_{c1})}{\BR(\x\to\pp\jpsi)}$
                & $0.88^{+0.33}_{-0.27}\pm0.10$ & BESIII~\cite{Bs3}    \\\hline
			 \specialrule{0em}{1.5pt}{1.5pt}
23    & $\frac{\BR(\x\to \gamma\psip)}{\BR(\x\to\gamma\jpsi)}$
                         & $2.46\pm0.64\pm0.29$ & LHCb~\cite{Lb1}      \\\hline
			 \specialrule{0em}{1.5pt}{1.5pt}
      & $B^+\to \x K^+$  & ($\times10^{-4}$)    &                      \\\hline
24    &                  & $2.1\pm 0.6\pm 0.3$  & BaBar~\cite{Ba2019d} \\
25    &                  & $1.2\pm 1.1\pm 0.1$  & Belle~\cite{Be2018d} \\
\hline\hline
\end{tabular}
\end{table*}

In the above definition, we assumed that all the measurements
follow Gaussian distribution, the statistical uncertainties
of different measurements are uncorrelated,
and possible correlation among the systematic 
uncertainties of different measurements in an experiment 
is neglected since the statistical uncertainties 
are dominant for most of the measurements. We also
assume that there is no correlation between different experiments.
 
The branching fractions 
$\BR(\x\to \pp\jpsi)$, $\BR(\x\to \ddstbn)$, 
$\BR(\x\to \gamma \jpsi)$, $\BR(\x\to \gamma\psp)$,
$\BR(\x\to \omega \jpsi)$, $\BR(\x\to \pi^0 \chi_{c1})$, 
$\BR(B^+\to \x K^+)$, and $\BR(B^0\to \x K^0)$ 
are free parameters in the fit. By minimizing 
$\chi^2(x)$ with {\sc minuit}~\cite{minuit}, the fitting results are 
obtained and listed in Table~\ref{tab:table2}. The fit
yields $\chi^2/{\rm ndf}$=25.2/17 where ${\rm ndf}$ 
represents the number of degrees of freedom.
The correlation coefficients between the fit parameters are 
shown in Table~\ref{tab:table3}.

\begin{table*}[htbp]
\caption{\label{tab:table2}
The fitting results of the absolute branching fractions 
of the $\x$ decays and $B\to \x K$ decays. The branching 
fraction of $\x$ decays into unknown modes is calculated
from the fit results.}
\begin{tabular}{cll}
\hline\hline
Parameter index & Decay mode & Branching fraction                     \\
\hline
1 & $\x\to \pp\jpsi$             & $(4.1^{+1.9}_{-1.1})\%$            \\
2 & $\x\to \ddstbn$              & $(52.4^{+25.3}_{-14.3})\%$         \\
3 & $\x\to \gamma\jpsi$          & $(1.1^{+0.6}_{-0.3})\%$            \\
4 & $\x\to \gamma \psp$          & $(2.4^{+1.3}_{-0.8})\%$            \\
5 & $\x\to \pi^0 \chi_{c1}$      & $(3.6^{+2.2}_{-1.6})\%$            \\
6 & $\x\to \omega \jpsi$         & $(4.4^{+2.3}_{-1.3})\%$            \\
7 & $B^+\to \x K^+$              & $(1.9\pm 0.6)\times 10^{-4}$       \\
8 & $B^0\to \x K^0$              & $(1.1^{+0.5}_{-0.4})\times 10^{-4}$\\\hline
  & $\x\to {\rm unknown}$        & $(31.9_{-31.5}^{+18.1})\%$         \\
\hline\hline
\end{tabular}
\end{table*}

\begin{table*}[htbp]
\caption{\label{tab:table3}
Correlation coefficients of the fit parameters 
listed in Table~\ref{tab:table2}.}
\begin{tabular}{ccccccccc}
\hline\hline
Parameter index &1&2&3&4&5&6&7&8\\
\hline
1&1&0.87&0.84&0.75&0.64&0.79&-0.95&-0.87\\
2&&1&0.79&0.71&0.56&0.74&-0.90&-0.77\\
3&&&1&0.78&0.54&0.73&-0.88&-0.78\\
4&&&&1&0.49&0.65&-0.79&-0.69\\
5&&&&&1&0.51&-0.61&-0.56\\
6&&&&&&1&-0.82&-0.72\\
7&&&&&&&1&0.84\\
\hline\hline
\end{tabular}
\end{table*}

From the fitting results we can see that 
$\ddstbn$ is the dominant decay mode of the $\x$
with a decay rate of $(52.4^{+25.3}_{-14.3})\%$, 
and all the other modes with charmonium have branching 
fraction of a few percent.
With all of the known branching fractions,
the fraction of $\x$ decays which is not yet observed in experiment 
is determined to be $(31.9_{-31.5}^{+18.1})\%$, so the
search for new decay modes of the $\x$ is still an important task.
As free parameters in the fitting, the production rates
of the $\x$ in neutral and charged $B$-meson decays are 
obtained as listed in Table~\ref{tab:table2}.

BESIII measured the product 
   $\sigma(\EE\to \gamma\x)\BR(\x\to \pp\jpsi)$
at center-of-mass energies from 4 to 4.6~GeV~\cite{Bs2},
with known $\BR(\x\to \pp\jpsi)$, the production cross section
of $\EE\to \gamma\x$ can be obtained. The peak cross section
of $\EE\to \gamma \x$ is found to be $(5.5^{+2.8}_{-3.6})$~pb 
at the center-of-mass energy 4.226~GeV. This is obtained by
sampling both numerator and denominator with the consideration 
of their asymmetric uncertainties, the cross section is 
about an order of magnitude smaller than that 
of $\EE\to \pp\jpsi$ at the same energy~\cite{pipijpsi_lineshape}.

In summary, we obtain the absolute branching fractions 
of the $\x$ decays into six modes by globally fitting the 
measurements provided by Belle, BaBar, BESIII, and LHCb 
experiments. The branching fraction of $\x\to \pp\jpsi$ is determined 
to be $(4.1^{+1.9}_{-1.1})\%$ which can be used in many
measurements where $\x\to \pp\jpsi$ is reconstructed and can
supply critical input in understanding the 
nature of the $\x$~\cite{ortega}. 
This branching fraction is in good agreement with 
early estimations in Refs.~\cite{ycz_pic2009} 
and~\cite{Esposito:2014rxa}. By combining the branching 
fractions of the observed modes, 
we obtain the fraction of the unknown decays of the $\x$
as $(31.9_{-31.5}^{+18.1})\%$, which calls for more experimental 
efforts in the study of the $\x$ decays. 

\acknowledgments 
The results were first presented at the Workshop
``Exotic Hadrons: Theory and Experiment at Lepton and Hadron Colliders'',
the authors thank the organizers, Prof. Luciano Maiani and Prof. Wei Wang,
for their hospitality. This work is supported in part 
by LiaoNing Revitalization Talents 
Program No. XLYC1807135; Scientific Research 
Foundation of LiaoNing Normal University No. BS2018L005;  
National Natural Science
Foundation of China (NSFC) under contract Nos. 11835012 and 11521505; 
Key Research Program of Frontier Sciences, CAS, 
Grant No. QYZDJ-SSW-SLH011; and the CAS Center for
Excellence in Particle Physics (CCEPP).

{\it Note added.}---After this paper was submitted, a paper titled
``Branching Fractions of the $\x$" by Braaten, He, and  Ingles~\cite{braaten} appeared.
Instead of treating the $\x$ as a single structure which has
universal properties in different production and decay mechanisms
as in our paper, these authors considered possible contributions
of $\ddstbn$ threshold effect and a bound state below the threshold
or a virtual state in $B^+\to K^+\x$ decay, and reported estimations
of the branching fractions of a bound state of charm mesons
(part of the $\x$ structure).
Future measurements of the line shapes of the $\x$ structures in
$B^+\to K^+\x$, $B^0\to K^0\x$, and $\EE\to \gamma\x$ may
check these above assumptions and reveal the nature of the $\x$.

\end{document}